\documentstyle[12pt]{article}
\begin{document}
\baselineskip 20pt plus 2pt

\setlength {\oddsidemargin}{0.5cm}
\setlength {\textheight}{21.0cm}
\setlength {\topmargin}{ 0.cm}

\hsize=15.5 true cm
\hspace{10cm} RU-94-28

\hspace{10cm} Technion PH-12-94

\vspace{2.5cm}
\begin{center}
{\bf A GENERALIZED ENSEMBLE OF RANDOM MATRICES~~ }\\

\vspace{1.0cm}
Moshe Moshe, \ \ \ Herbert Neuberger$^{(a)}$,\ \ \ Boris Shapiro\\

\vspace{1.0cm}
Department of Physics\\
Technion - Israel Institute of Technology\\
32000 Haifa, Israel.

\vspace{1.0cm}
{\bf Abstract}
\end{center}
\vspace{0.50cm}

\noindent  A random matrix ensemble incorporating both GUE
and Poisson
level statistics while respecting $U(N)$
invariance is proposed and shown
to be equivalent to a system of noninteracting,
confined, one dimensional
fermions at finite temperature.
\vspace{1.5 cm}

\vspace{0.2cm}

\noindent\rule{10cm}{0.02cm}

\noindent (a) Permanent address:
Rutgers University, Department of Physics,
Piscataway NJ 08855, U.S.A.
\pagebreak

Wigner's$^{[1]}$ suggestion to model
the level statistics of complex quantum
systems by random matrix models of very
large dimensions is well known.  In
applications as diverse as  two dimensional
gravity$^{[2]}$, topological field
theory$^{[3]}$ quantum chaos$^{[4]}$ and
mesoscopic physics$^{[5]}$ different
kinds of universal features emerge, all
represented by particular types
of random matrix ensembles.  The analysis
of these models seems to lead
one to classical$^{[2,3]}$ and quantum
integrable systems$^{[6]}$ of very
special mathematical structure and many
unexpected relations make
their appearance$^{[7]}$.

Random matrix models exhibit a large degree
of robustness$^{[8]}$ and
the search for different kinds of universality
has produced a wealth of
 results in two dimensional quantum gravity$^{[2]}$
 and in topological field
theory$^{[3]}$.
In quantum chaos and mesoscopic physics there are
physical reasons to
expect more universality classes than presently
known$^{[9-12]}$.
This letter takes a small step in this direction.

For reasons of concreteness and mathematical
simplicity, we restrict our
attention to a disordered system
without time-reversal symmetry (e.g. in
the presence of a magnetic field)
and ask whether random matrix models
can account for known departures from
GUE (Gaussian Unitary Ensemble) level
statistics$^{[10-13]}$.  On the
one hand the class of matrix
models with GUE level statistics has
been shown recently to be very
wide$^{[8,14]}$ and also to agree well
with the properties
in the energy region
of extended states$^{[15]}$, but, on the
other hand,  disordered
systems have
energy regions where the states are
localized and the level statistics
is Poissonian.  To be sure, even
localized states, once sufficiently
close energetically, eventually
repel each other because the
exponentially small overlap can be
outweighted by exceedingly small
energy denominators.  It is the typical
scale of energy differences
needed for level repulsion to be felt
that changes dramatically in
transition from extended to localized regimes.

Recently, an attempt to incorporate
both statistics in a random matrix
model was made$^{[12]}$: The basic
idea was that in the localized
regime the Hamiltonian prefers the
site basis and hence the random matrix
model was made to break the $U(N)$
invariance which represents the lack
of basis preference in the GUE ensemble.
For sufficiently strong breaking,
approximate Poisson statistics
could be obtained.  The explicit breaking
of $U(N)$  invariance makes the
mathematical treatment and the study of
robustness difficult.
While the models we are discussing
cannot be fully accurate representations
of the mesoscopic physics since we
will not introduce any explicit dependence
on dimension, we feel that exploring
the possibility for constructing such
an enlarged universality class is of
more general interest and
holds the potential for wider applicability
than just to mesoscopic physics.

Our main new observation is that basis preference,
while needed, does not
necessarily force the abolishment
of $U(N)$ symmetry.  In a way vaguely reminiscent
of spontaneous symmetry breaking, we argue
that what is important is the
preference for {\it {\bf some}}
basis but it is immaterial which basis this is.
Averaging over all preferred bases we recover
$U(N)$ invariance and still
manage to evade GUE statistics if a certain
parameter is made strong enough.

Imagine that the preferred basis is represented
by an $N\times N$ unitary
matrix $V$. This representation has $N$ redundant
phases and it is
therefore convenient to consider instead
the unitary matrix $U = V D V^\dagger$ where
$D_{ij} = \delta_{ij}e^{i\theta_i}$,
{}~(~$|\theta_i| \leq \pi$,
$i=1,2,\dots N$~).
angles $\theta_i$ but
For $U$ to define uniquely a preferred basis
we need the
$e^{i\theta_j}$'s to be all different.
In order to ensure this upon averaging  we would
like the  weight to contain sufficient
repulsion between the
$e^{i\theta_j}$'s.
Let the Hamiltonian be represented by
a hermitian matrix $H$.  For
fixed $U$ we consider the unnormalized
distribution of $H$  given by:
\begin{eqnarray}
P_U(H)d^{N^2}H~~
\propto  ~~e^{-TrH^2}
e^{-bTr([U,H][U,H]^\dagger)}{d^{N^2}}H     
\end{eqnarray}
Here,
$d^{N^2}H= \prod^N_{i=1}
dH_{ii}\prod_{i>j} dRe(H_{ij})dIm(H_{ij})$
and $b>0$.  The $b$-dependent term
tries to align $H$ with $U$ and hence
prefers the basis $V$.  We now average
over $U$ with the invariant $U(N)$
Haar measure $dU$.  This induces
enough repulsion between the eigenvalues of
$U$ $^{[1,16]}$ to ensure that a unique
preferred basis almost always
exists.  Our ensemble therefore is:
\begin{eqnarray}
P(H)d^{N^2}H = C^\prime (N,b)~
e^{-(2b+1)TrH^2}
[~\int dU
e^{2bTr(UHU^\dagger H)}~]d^{N^2}H     
\end{eqnarray}
$C^\prime (N,b)$ is a normalization constant.

The integral over $U$ can be done$^{[17]}$
and, going over
to the marginal distribution for the
eigenvalues of $H$ , $x_i$, we find:
\begin{eqnarray}
P(x_1,x_2,\cdots x_N)
\prod_i dx_i
=&&C(N,b)e^{-\sum_ix_i^2}~
{\cal D}et_{ij} \left(e^{-b(x_i-x_j)^2}\right)
\prod_i dx_i \nonumber \\
=&&C(N,b){\cal D}et_{ij}\left(
e^{-(b+\frac{1}{2})(x_i^2+x_j^2)
+2bx_ix_j} \right)\prod_i dx_i
\end{eqnarray}
 $P(x_1,x_2,\cdots x_N)$ in
 Eq.(3) coincides$^{[18-20]}$
with the diagonal element in the
``coordinate representation $(x)$"
 of the density matrix of $N$ non-interacting
fermions of mass $m$, in one dimension,
moving in a confining harmonic
potential $V(x)=\frac{1}{2}m\omega^2x^2$
at finite temperature,
\begin{eqnarray}
P(x_1,x_2,\cdots x_N)=&&{\cal N}
{\cal D}et_{ij}\left(<x_i|
exp\{-\beta(\frac{\hat p^2}{2m}
+\frac{m\omega^2\hat q^2}{2})\}
|x_j> \right) \nonumber \\
=&&{\cal N}{\cal D}et_{ij}\left(
exp\{-\frac{m\omega}{2\hbar}
coth(\hbar\omega\beta)(x_i^2 + x_j^2)
+ \frac{m\omega}{\hbar}
\frac{x_ix_j}{
sinh(\hbar\omega\beta)} \} \right)~~~~~~,
\end{eqnarray}
where ${\cal N}$ is a normalization factor,
$\beta\hbar\omega={\rm arcosh}(1+\frac{1}{2b})$ and
the length unit is fixed by
$\frac{m\omega}{\hbar}=\sqrt{1+4b}$.
This means that the mass and
temperature are related by
$\frac{m\omega}{\hbar} = {\rm coth}
\frac{\hbar\omega\beta}{2}$.
The GUE ensemble is obtained in the limit
$\hbar\omega\beta
\rightarrow \infty, \frac{m\omega}{\hbar}
\rightarrow 1$
(implied by $b \rightarrow 0$, GUE
statistics in Eq.(3) ) and
complete disappearance of level
repulsion is obtained in the limit
 $\hbar\omega\beta
\rightarrow 0, \frac{m\omega}
{\hbar}\rightarrow \infty$
($b \rightarrow \infty$, Poisson
statistics in Eq. (3) ).

When $N$ is large it is much more
convenient to analyze the grand canonical version
of the ensemble (3).
Indexing the one particle states by $\alpha$,
denoting their energies by
$\epsilon_\alpha$ and the corresponding wave
functions by $\psi_\alpha(x)$ one
immediately can write down the
expression for the local density $\bar{n}(x)$:
\begin{eqnarray}
\bar{n}(x) =
\sum_\alpha
\frac{|\psi_\alpha(x)|^2}
{e^{\beta(\epsilon_\alpha-\mu)}+1}     
\end{eqnarray}
The chemical potential $\mu$ is fixed by $N$ via
$N = \sum_\alpha
[e^{\beta
(\epsilon_\alpha-\mu)}+1]^{-1} $ \ .      

For large $N$ one can use semiclassical forms for
$\psi_\alpha(x)$ as long as $x$ is
away from the edges of the support of
$\bar{n}(x)$ and replace the sum
over $\alpha$ by an integral.
A simple computation gives then:
\begin{eqnarray}
\bar{n}(x) =
\int^\infty_{-\infty}
\frac{dp}{2\pi\hbar}
{}~~\frac{1}
{1+[{e^{\beta\hbar\omega N}-1}]^{-1}
e^{\beta(\frac{1}{2m}p^2
+\frac{1}{2}m\omega^2x^2)} }
\end{eqnarray}
Eq.~(6) already exhibits both limiting
situations, of GUE and of
Poisson statistics: For $b$ growing with
$N$ slower than $N^2$,
$\beta N \rightarrow \infty$, the gas
becomes completely
degenerate and one obtains the GUE semicircle law:
\begin{eqnarray}
\bar{n}(x) =
\frac{\sqrt{4b+1}}{\pi}
\left[ \frac{2N}{\sqrt{2b + 1}} - x^2
\right]^{\frac{1}{2}}   
\end{eqnarray}
For b growing with $N$ faster than $N^2$,
$\beta N \rightarrow 0$, a gaussian distribution
is obtained:
\begin{eqnarray}
\bar{n}(x) =
\frac{N}{\sqrt{\pi}}~{e^{-x^2}}          
\end{eqnarray}
Correlations can also be easily calculated.
Consider the 2-point
connected correlation:
\begin{eqnarray}
\overline{\delta n(x)\delta(n)(x^\prime)} =
 -|\sum_\alpha \bar{n}_\alpha
\psi^*_\alpha (x)\psi_\alpha(x^\prime)|^2 \
 ~~~~~~~~~~~~~~~x \neq x^\prime          
\end{eqnarray}
with $\bar{n}_\alpha =
[e^{\beta(\epsilon_\alpha-\mu)} + 1]^{-1}$.
Writing
$x = \bar{x}+\frac{1}{2}\Delta x$,
$x^\prime = \bar{x}-\frac{1}{2}\Delta x$
with $0<|\Delta x|<< \bar{n}
(\frac{d\bar{n}}{dx})^{-1}$ we derive
for $\beta N \rightarrow \infty$:
\begin{eqnarray}
\frac{\overline{\delta n(x)\delta
n(x^\prime)}}{(\bar{n}(\bar{x}))^2}
= -
\left[ \frac{\sin(\pi\bar{n}(\bar{x})
\Delta x)}{\pi\bar{n}(\bar{x})\Delta x}
\right]^2
\end{eqnarray}
and for $\beta N \rightarrow 0$:
\begin{eqnarray}
\frac{\overline{\delta n(x)\delta
n(x^\prime)}}{(\bar{n}(\bar{x}))^2}
= -
e^{-2b(\Delta x)^2} ~~~\rightarrow 0      
\end{eqnarray}
The last two equations show that the
model incorporates GUE
and Poisson statistics.

Consider now the intermediate case when, in the
$N \rightarrow \infty$
limit,
$\beta N \rightarrow \mbox{const} \ \equiv A$.
If the constant $A$ is small, the
average density of levels
$\bar{n}(x)$
and the correlations are approximately the same as for
$\beta N \rightarrow 0$.
The case
$A >> 1$
is more interesting:
$\bar{n}(x)$
is a complicated function which,
for small $x$, is proportional
to $\sqrt{A-x^2}$
and, for large $x$, is proportional to
$\exp (-x^2)$
(``small'' and ``large'' here mean, respectively,
$(A-x^2) >> 1$
and
$-(A-x^2) >> 1$).
The correlation also changes  when
one moves from the
center of the band
towards the edges: \ at the edges
(i.e.
$x >> A$)
the levels are essentially uncorrelated,
whereas at the band center
$(x \approx 0)$
the correlation function is given by
\begin{equation}
\overline{ \frac{\delta n (x)
\delta n (x')}{(\bar{n} (x))^2} }
=
-\left(  \frac{\sin \pi u}{\pi u} \right)^2
\left[ \frac{\pi^2u/2A}{\sinh
(\pi^2 u/2A)} \right]^2 \ ,
\end{equation}
where
$u \equiv \bar{n} (x) \Delta x$
is the energy separation measured in units
of the averaged
level spacings. While a sequence of roughly $A$ levels
obeys GUE statistics, the correlations
between more distant
levels rapidly diminish.  This is in
sharp contrast with the
case
$\beta N \rightarrow \infty$
($\beta N \rightarrow 0$)
where, in the limit
$N \rightarrow \infty$,
an arbitrary long sequence of levels
obey the GUE (Poisson)
statistics.
In this sense the parameter
$\beta N$
of our model can be identified with
the dimensionless
conductance $g$.
$g \rightarrow \infty$
corresponds to a metallic behavior,
$g \rightarrow 0$
corresponds to an insulator and
$g \rightarrow const$
is identified with a mobility edge
situation, where a third
universal statistics, intermediate
between the GUE and Poisson,
is observed$^{10-12}$. In  that latter case
$\overline{\Delta N^2_{\Gamma}} \approx
\bar{N}_{\Gamma}/2A$
where
$\overline{\Delta N^2_{\Gamma}}$
describes fluctuations within an energy
interval $\Gamma$ with
an average number of levels equal to
$\bar{N}_{\Gamma}$
$(\bar{N}_{\Gamma} >> A$
is assumed). The linear relation between
$\overline{\Delta N^2_{\Gamma}}$
and
$\bar{N}_{\Gamma}$
is a consequence of the exponential decay
of correlations in
Eq. (12).
In ref. 11 it was argued that a linear
dependence is ruled out in the
universal regime by a sum-rule expressing
the conservation of the
total number of levels, $N$. In our
analysis of the matrix model
we found that in the large $N$ limit
it is correct to pass from
the canonical description - fixed number
of levels or fermions -
to the grand canonical one - arbitrary
number of levels in the
presence of a ``chemical potential''.
Therefore, the overall level
number conservation cannot
affect universal features of the statistics
of finite sequences of
levels in the thermodynamic limit.

A guess for the enlarged universality class
of these models
is:~  add more $U(N)$ invariant function
of $U$ and $H$ to the
exponent in Eq.~(1) and average over $U$.
While strict $U(N)$ invariance
is useful, models with explicit breaking$^{[12]}$
are not excluded and can easily turn out to
be controlled by $U(N)$-symmetric fixed points.
In this context it is
amusing to note that if, in a reversal of roles,
 one integrates over $H$ in (1) and views
the resulting distribution of $U$ as a
generalized circular Dyson-type
ensemble$^{[16]}$ one recovers a ``circular''
version of the eigenvalue
distribution arrived at in Ref.~[12]:
\begin{eqnarray}
P_\theta(\theta_1,\dots\theta_N)
\prod^N_{i=1} d\theta_i =
C^{\prime\prime}(N,b)
\prod_{i>j}
\left[ \frac{\sin^2
(\frac{\theta_i-\theta_j}{2})}
{1+4b\sin^2
(\frac{\theta_i-\theta_j}{2})}
\right] \prod^N_{i=1} d\theta_i  
\end{eqnarray}
For numerical purposes one may wish
 to stick to the canonical ensemble;
 expressions for $\bar{n}(x)$
and for
$\frac{\overline{\delta n(x)\delta
n(x^\prime)}}{(\bar{n}(x))^2}$
can be obtained in terms of infinite
sums that converge fast as long as
$\beta N$ does not grow to infinity
$^{[20,21]}$.  For example, the
exact formula for $\bar{n}(x)$, in Eq.~(5), is:
\begin{eqnarray}
\bar{n}(x) =
\sum^\infty_{\alpha = 1} |\psi_\alpha(x)|^2
\left\{ \sum^\infty_{r=0} (-1)^r
q^{(N-\alpha)r+\frac{1}{2}r(r+1)}
\prod^{N+r}_{\ell=N+1} (1-q^\ell)^{-1} \right\}   
\end{eqnarray}
where
$\epsilon_\alpha = \hbar\omega
(\alpha-\frac{1}{2})$,
$q = e^{-\beta\hbar\omega} =
\left[ 1+\frac{1}{2b} + \sqrt{\frac{1}{b}
+\frac{1}{4b^2} } \right]^{-1}$, and
$\prod^{N}_{\ell=N+1} (1-q^\ell)^{-1} \equiv 1$.

Eq.~(2) can be
viewed as a degenerate case, in one
space-time dimension, of
``induced QCD'' models$^{[22]}$.
Generalizations of the ensemble discussed
here to the orthogonal or
symplectic case are likely to produce
technical complications
similar to the ones encountered in the
study of random unoriented two
dimensional manifolds with or without matter$^{[23]}$.

In summary, the random matrix model
proposed in eqs. (2,3)
 is an extension of the standard
 gaussian unitary ensemble
 with a level distribution that
 interpolates between
 Wigner--Dyson and Poisson statistics.
 $U(N)$ invariance
 is preserved and the model is equivalent
 to a system
 of one dimensional fermions in a harmonic
 potential
 at finite temperature. Possible applications
  are to modeling of level statistics
  in quantum
 chaos, mesoscopic physics and disordered
 electronic
 systems. For example, in the latter case,
 the proposed ensemble can be driven from
 metallic to insulator behavior by varying
 the parameter $b$.

\noindent{\bf Acknowledgments}

We would like to express our thanks to
U. Sivan for  a clarifying
discussion and to M. Staudacher
for help with references.
H.N. would like to thank the
Institute for Theoretical Physics at the
Technion for the warm
hospitality extended while this work was
performed.  This research was
supported in part by the BSF under grant
\# 92-00201 (090-755) (MM and HN),
by the DOE under
grant \# DE-FG05-90ER40559 (HN) and by the Technion
V.P.R Fund (MM and BS).

\pagebreak
\baselineskip 14pt plus 2pt

\noindent{\bf References}
\begin{enumerate}
\item E.P. Wigner, Proc.\
Cambridge Philos.\ Soc.\ {\bf 47}, 790 (1951).\\
M.L. Mehta, ``Random Matrices'',
2nd ed. (Academic Press, New York, 1991).
\item E. Br\'{e}zin, V. Kazakov,
Phys.\ Lett.\ {\bf 236b}. 2125 (1990)\\
M.R. Douglas, S.H. Shenker, Nucl.\
Phys. {\bf B335}, 635 (1990) \\
D.J. Gross, A.A. Migdal, Phys.\ Rev.\
Lett. {\bf 64}, 27 (1990).
\item E. Witten, Nucl.\ Phys.\ {\bf B340},
281 (1990).
\item M.C. Gutzwiller, ``Chaos in Classical
and Quantum Mechanics''\\
(Springer Verlag, N. Y. , 1990) \\
T. M. Seligman, J. J. M. Verbaarschot and
M. R. Zirnbauer, Phys. Rev. Lett
{\bf 53}, 215 (1984); Phys. Lett. {\bf 108A},
183 (1985)
\item L.P. Gorkov, G.M. Eliashberg,
Sov.\ Phys.\ JETP {\bf 21}, 940 (1965).\\
Y. Imry, Europhys.
\ Lett.\ {\bf 1}, 249 (1986). \\
B.L. Altshuler, B.I. Shklovskii,
Sov.\ Phys.\ JETP {\bf 64}, 127 (1986).
\item B.D. Simons, P.A. Lee, B.L. Altshuler,
Phys.\ Rev.\ Lett. {\bf 70}, 4122 (1993).
\item B.D. Simons, P.A. Lee, B.L. Altshuler,
Phys.\ Rev.\ Lett. {\bf 72}, 64 (1994).
\item E. Br\'{e}zin, A. Zee, Nucl.\ Phys.
{\bf B402} (FS), 613 (1993).
\item T. M. Seligman, J. J. M. Verbaarschot
and M. R. Zirnbauer,
Phys.\ Rev.\ Lett. {\bf 53}, 215 (1984),
Phys.\ Lett. {\bf 108A}, 183 (1985).
\item  B.I. Shklovskii, B. Shapiro, B.R. Sears,
P. Lambrianides and H.B. Shore,
Phys.\ Rev.\ {\bf B47}, 11487 (1993)
\item V.E. Kravtsov, I.V. Lerner,
B.L. Altshuler, A.G. Aronov, Phys.\ Rev.\ Lett.
{\bf 72}, 888 (1994).
\item J.-L. Pichard, B. Shapiro,
Saclay and Technion preprint, 1993.
{}~~~To be published in Journal de Physique
1, May 1994
\item Y.V. Fyodorov and A.D. Mirlin,
Phys.\ Rev.\ Lett.\ {\bf 69}, 1093
(1992) and Phys.\ Rev.\ Lett.\ {\bf 71}, 412 (1993).
 \item E. Br\'{e}zin and A. Zee, ENS preprint (1993).
\item B.D. Simons,  B.L. Altshuler, Phys.\ Rev.\ Lett.
{\bf 70}, 4063 (1993).\\
A. Szafer, B.L. Altshuler,
Phys.\ Rev.\ Lett.\ {\bf 70} 587 (1993).
\item F.J. Dyson, J.\ Math.\ Phys.
{\bf 3}, 1199 (1962).
\item C. Itzykson, J.B. Zuber, J.\ Math.\ Phys
{\bf 21}, 411 (1980).
\item E. Br\'{e}zin, C. Itzykson, G. Parisi
and J.B. Zuber,
Comm. Math. Phys. {\bf 59}, 35, (1978)
\item G. Parisi, Phys.\ Lett.\
{\bf 238B}, 213 (1990).
\item D. Boulatov, V. Kazakov, Nucl.
Phys. B (Proc. Suppl.) {\bf 25A}, 38 (1992).
\item M. Caselle, A. D'Adda,
S. Panzeri, Phys.\ Lett.\ {\bf 293B},
161 (1992).
\item V.A. Kazakov, A.A. Migdal,
Nucl.\ Phys.\ {\bf B397}, 214 (1993).
\item E. Br\'{e}zin, H. Neuberger,
Nucl.\ Phys.\ {\bf B350}, 513 (1990).\\
H. Neuberger, Phys.\ Lett.\ {\bf 257B}, 45 (1991).
\end{enumerate}
\end{document}